\newcommand{\HI}{\mbox{H\,{\sc i}}}
\newcommand{\HII}{\mbox{H\,{\sc ii}}}

\documentclass{emulateapj}

\usepackage{natbib}
\usepackage{color}
\usepackage{mathrsfs}

\slugcomment{Not to appear in Nonlearned J., 45.}

\shorttitle{KS relation variety and SF cloud fraction}
\shortauthors{Morokuma-Matsui et al.}

\begin{document}


\title{Kennicutt-Schmidt relation variety and star-forming cloud fraction}

\author{Kana Morokuma-Matsui\altaffilmark{1} and Kazuyuki Muraoka\altaffilmark{2}}
\email{kana.matsui@nao.ac.jp}


\altaffiltext{1}{Chile Observatory, National Astronomical Observatory of Japan,
    2-21-1 Osawa, Mitaka-shi, Tokyo 181-8588, Japan}
\altaffiltext{2}{Department of Physical Science, Osaka Prefecture University, 1-1 Gakuen-cho, Naka-ku, Sakai, Osaka 599-8531, Japan}


\begin{abstract}
The observationally derived Kennicutt-Schmidt (KS) relation slopes differ from study to study, ranging from sub-linear to super-linear.
We investigate the KS-relation variety (slope and normalization) as a function of integrated intensity ratio, $R_{31}=$~CO($J=3-2$)/CO($J=1-0$) using spatially resolved CO($J=1-0$), CO($J=3-2$), $\HI$, H$\alpha$ and 24~$\mu$m data of three nearby spiral galaxies (NGC~3627, NGC~5055 and M~83).
We find that
(1) the slopes for each subsample with a fixed $R_{31}$ are shallower but the slope for all datasets combined becomes steeper,
(2) normalizations for high $R_{31}$ subsamples tend to be high,
(3) $R_{31}$ correlates with star-formation efficiency, thus the KS relation depends on the distribution in $R_{31}$-$\Sigma_{\rm gas}$ space of the samples: no $\Sigma_{\rm gas}$ dependence of $R_{31}$ results in a linear slope of the KS relation whereas a positive correlation between $\Sigma_{\rm gas}$ and $R_{31}$ results in a super-linear slope of the KS relation, and
(4) $R_{31}$-$\Sigma_{\rm gas}$ distributions are different from galaxy to galaxy and within a galaxy: galaxies with prominent galactic structure tend to have large $R_{31}$ and $\Sigma_{\rm gas}$.
Our results suggest that the formation efficiency of star-forming cloud from molecular gas is different among galaxies as well as within a galaxy and is one of the key factors inducing the variety in galactic KS relation.

\end{abstract}


\section{Introduction}
\label{sec:introduction}

Star formation is one of the essential processes of galaxy evolution.
However, the physics of star formation is not fully understood because one must consider a large dynamic range of spatial scale as well as time scale, from kilo-parsec to 0.1~parsec and from Giga-year to Mega-year, respectively.

Schmidt (1959) first proposed a star formation model by assuming that the volume density of the star-formation rate ($\dot{\rho}_\star$) varies with a power $n$ of the volume density of gas ($\rho_{\rm gas}$) as $\dot{\rho}_\star \propto \rho_{\rm gas}^n$.
Thirty years later, Kennicutt (1989) observationally showed a strong relation between the surface densities of star formation rate (SFR), $\Sigma_{\rm SFR}$, and cold gas mass, $\Sigma_{\rm gas}$, inferred from the $\HI$ and CO($J=1-0$) (hereafter, CO(1-0)) luminosities of nearby galaxies: $\Sigma_{\rm SFR} \propto \Sigma_{\rm gas}^{1.4}$.
In a more general form that incorporates a power of $N$, the relation $\Sigma_{\rm SFR} \propto \Sigma_{\rm gas}^N$ is referred to as the Kennicutt-Schmidt (KS) relation.

Since the proposition of the KS relation, various studies have reported various normalizations and slopes of the KS relation mainly based on molecular gas (hereafter molecular KS relation), -- i.e., $\Sigma_{\rm gas} \propto \Sigma_{\rm mol}^{N_{\rm mol}}$, where $\Sigma_{\rm mol}$ is molecular gas surface density.
For example, higher normalizations are suggested for starbursts \citep[e.g.,][]{Daddi:2010lr} and smaller normalizations for early-type galaxies \citep[e.g.,][]{Martig:2013fr} compared to normal disk galaxies.
Even for the normal disk galaxies, various slopes are reported including sub-linear \citep[e.g.,][]{Shetty:2013yb}, linear \citep[e.g.,][]{Bigiel:2008rt}, and super-linear \citep[e.g.,][]{Momose:2013yq}.
Many studies suggested that the variation is due to the differences in methodology such as SFR estimation (\citealt{Liu:2011kx,Momose:2013yq}, see also \citealt{Rahman:2011ys,Rahman:2012fr}), molecular gas mass estimation \citep{Momose:2013yq}, fitting method \citep[e.g.,][]{Blanc:2009vn,Verley:2010rt}, and spatial resolution of the data \citep{Kennicutt:2007fk,Thilker:2007qy,Onodera:2010uq,Momose:2010fj,Liu:2011kx,Calzetti:2012lr}.

Some studies suggested that the reported KS-relation variety is not only due to different methodologies but also genuine and intrinsic.
\cite{Lada:2012zr} explicitly showed that the Milky Way molecular clouds and nearby galaxies with the same dense core gas fraction follow the same linear molecular KS relation and that the normalizations tend to be high for systems with high dense core gas fraction.
This is based on the observational result that SFR correlates linearly with dense core mass (traced by HCN) in molecular clouds and galaxies \citep{Solomon:1992kl,Gao:2004kx,Gao:2004fj,Wu:2005ai}.
As a subsequent step, we must investigate the controlling factors of dense core gas fractions in various galactic environments and galaxies.
However, previous HCN observations were mainly conducted toward the centers of nearby galaxies, and only a few galaxies have been mapped recently, since the HCN emission is generally $10-30$ times weaker than CO (M82: \citealt{Kepley:2014fk}, Antennae: \citealt{Bigiel:2015qy}, M51: \citealt{Chen:2015lr,Bigiel:2016uq}).

It has been shown that line luminosities of the high-$J$ transitions of CO, the second most abundant molecule, also linearly correlate with SFR (or infrared luminosity) (\citealt{Liu:2015fj}, see also \citealt{Bayet:2009mi,Greve:2014rw}).
Especially, CO($J=3-2$) (hereafter, CO(3-2)) mapping observations were conducted toward a number of galaxies \citep[e.g.,][]{Wilson:2009kx} with observational indication that CO(3-2) correlates linearly with SFR over four orders of magnitudes \citep{Narayanan:2005qa,Komugi:2007xr,Bayet:2009mi,Iono:2009oz,Mao:2010kl,Greve:2014rw,Muraoka:2016lr}.
Although CO(3-2) emission is likely to be optically thick toward the dense cores, CO(3-2) traces dense ($\sim10^4$~cm$^{-3}$) and warm ($30-50$~K) gas, which is found around cores heated by star formation \citep[e.g.][]{Minamidani:2008wd,Kawamura:2009eu,Dempsey:2013fu,Miura:2014nx}.
In addition, CO(3-2) is considered to be an indicator of the early stage star formation, since it also traces outflows from protostars \citep[e.g.][]{van-Kempen:2006fk,Su:2007gf,Takahashi:2008ve}.

In this study, we investigate the relationship between the variety in the CO(3-2)/CO(1-0) ratio of nearby spiral galaxies and the resultant slopes of the KS relation.
We consider the CO(3-2)/CO(1-0) ratio as a measure of star-forming gas fraction.
The structure of this paper is as follows;
we first describe the observational data in Section~\ref{sec:data}.
Then we show the variety in the CO(3-2)/CO(1-0) ratios within a galaxy and among galaxies, and how the KS-relation slope depends on the CO(3-2)/CO(1-0) ratio in Section~\ref{sec:result}.
We propose ``hierarchical KS relation'' and discuss which hierarchical steps of interstellar medium (ISM) impose a variety in the KS relation and what controls the ISM hierarchy balance in disk galaxies in Section~\ref{sec:discussion}.

\section{Data}
\label{sec:data}

\begin{figure*}[t]
 \begin{minipage}{0.5\hsize}
  \begin{center}
   \includegraphics[width=70mm]{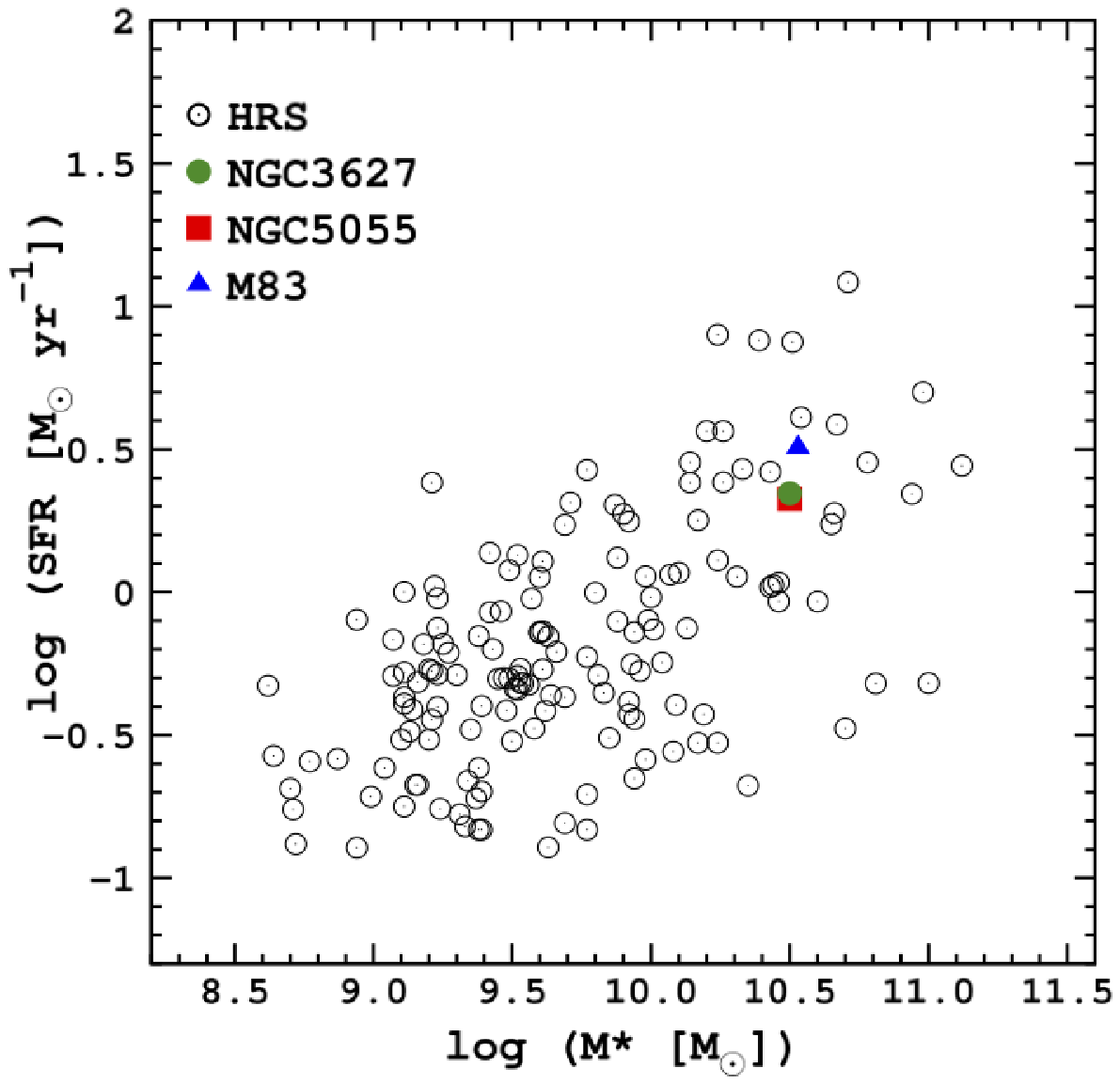}
  \end{center}
  \caption{Stellar mass and SFR relation of nearby galaxies including NGC~3627 (green circle), NGC~5055 (red square) and M~83 (blue triangle).
  Stellar masses based on 3.6~$\mu$m and SFRs of both NGC~3627 and NGC~5055 are retrieved from \cite{Leroy:2013qo} and \cite{Leroy:2008fb}, respectively.
  Stellar mass and SFR of M~83 are retrieved from \cite{Jarrett:2013bv}.
  The black-filled circle indicates the galaxies from Herschel Reference Survey \citep[HRS,][]{Boselli:2010lh} as a reference;
  stellar mass based on $i$-band luminosity and $g-i$ color from \cite{Cortese:2012kb}, and SFR based on FUV and 24~$\mu$m luminosity from \cite{Boselli:2015qf}.
  The sample galaxies are on the star-forming {\it main-sequence} and located closely.}
  \label{fig:MSFR}
 \end{minipage}
 \begin{minipage}{0.5\hsize}
  \begin{center}
   \includegraphics[width=85mm]{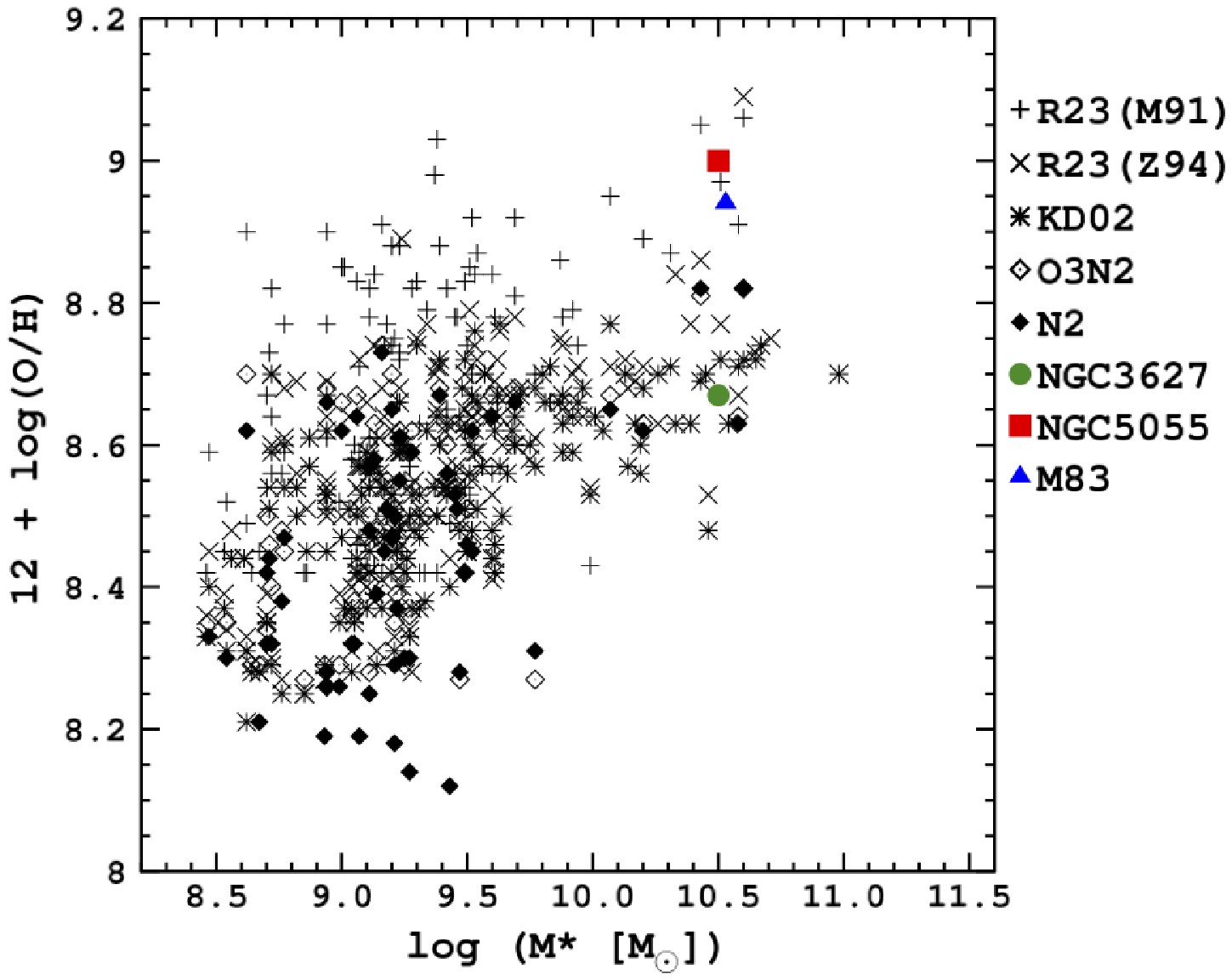}
  \end{center}
  \caption{Stellar mass and metallicity relation of nearby galaxies including NGC~3627 (green circle), NGC~5055 (red square) and M~83 (blue triangle).
  Metallicities based on R23 calibration with \cite{Kobulnicky:2004tw} (KK04) and \cite{Pilyugin:2005os} (PT05) of NGC~3627 and NGC~5055 are retrieved from \cite{Moustakas:2010jw}.
  Averaged values at the galactic center of PT05 and KK04 calibrations are adopted (Table.~9 of \citet{Moustakas:2010jw}).
  Direct measurement of metallicity at the center of M~83 based on electric temperature is retrieved from \cite{Bresolin:2005il}.
  Black symbols indicate HRS galaxies and the metallicities of them are retrieved from \cite{Hughes:2013ye};
  cross, R23 calibration based on \cite{McGaugh:1991ee}; x, R23 calibration based on \cite{Zaritsky:1994zt}; 
  asterisk, [Nii]$\lambda6584/$[Oii]$\lambda3727$ calibration based on \cite{Kewley:2002xw}; open diamond, O3N2 calibration based on \cite{Pettini:2004jl}; filled diamond, N2 calibration based on \cite{Pettini:2004jl}.
  Metallicities of NGC~5055 and M~83 seem higher than that of NGC~3627 even considering the difference in the calibration methods.}
  \label{fig:MZ}
 \end{minipage}
\end{figure*}

\begin{table*}
\begin{center}
\caption{Summary of the observational data used in this study \label{tab:obs}}
\begin{tabular}{lccccccccc}
\tableline\tableline
Galaxy & Type & Arm class$^{\rm a}$ & Dist. (ref.)$^{\rm b}$ & $i$ (ref.)$^{\rm c}$ & Resolution$^{\rm d}$ & CO(1-0) & CO(3-2) & $\HI$ & SFR\\
 & (RC3) & & (Mpc) & (deg) & (pc) & & & & \\
\tableline
NGC~3627 & SAB(s)b & 7 & 11.1 (1) & 52 (1) & 892 & NRO45 & JCMT & THINGS & SINGS\\
NGC~5055 & SA(rs)bc & 3 & 7.2 (2) & 61 (1) & 819 & NRO45 & JCMT & THINGS & LVL\\
M~83 & SAB(s)c & 9 & 4.5 (3) & 24 (2) & 575 & NRO45 & ASTE & THINGS & LVL\\
\tableline
\end{tabular}
\tablecomments{
$^{\rm a}$ 7 and 9 indicate a grand design spiral and 3 indicates a flocculent \citep{Elmegreen:1987kq}.\\
$^{\rm b}$ References of distance. 1: \cite{Saha:1999yu}, 2: \cite{Pierce:1994qf}, 3: \cite{Thim:2003sw}\\
$^{\rm c}$ References of inclination. 1: \cite{Kuno:2007uq}, 2: \cite{Comte:1981kx}\\
$^{\rm d}$ $i$-corrected resolution.
}
\end{center}
\end{table*}

A summary of the observational data used in this study is given in Table~\ref{tab:obs}.
All three galaxies are nearby late-type galaxies with grand design (NGC~3627 and M~83) or flocculent (NGC~5055) spiral structures \citep{Elmegreen:1987kq}.
They belong to a similar population of galaxies in terms of stellar mass and SFR plane (Figure~\ref{fig:MSFR}) but NGC~3627 seems to have lower metallicity than the other two galaxies (Figure~\ref{fig:MZ}).
We use archival data to estimate gas and SFR surface densities.
The spatial resolutions of the data are 15~arcsec for NGC~3627 (700~pc at the distance of the galaxy) and NGC~5055 (570~pc), and 25~arcsec for M~83 (550~pc).
The inclination corrected spatial resolutions are 892~pc for NGC~3627, 819~pc for NGC~5055, and 575~pc for M~83, which are larger than the boundary spatial resolutions at which the KS relation breaks \citep[$80-250$~pc,][]{Momose:2010fj,Verley:2010rt,Onodera:2010uq}.
In the following subsections, we briefly summarize the data and derivations of the physical values such as total cold gas mass and SFR from the observed quantities.

\subsection{Gas surface density and CO(3-2)/CO(1-0) ratio}

The estimation of gas surface density ($\Sigma_{\rm gas}$) requires data on the H$_2$ and $\HI$ column densities.
Here, the estimate of $\Sigma_{\rm gas}$ takes into account He content (multiplied with a factor of 1.36, based on the (proto-) Solar abundance, e.g., \citealt{Asplund:2009fh}).
CO(1-0) data for our sample of galaxies were obtained with the 45-m telescope as a part of Nobeyama CO Atlas survey \citep{Kuno:2007uq}.
We adopt a CO-to-H$_2$ conversion factor of $X_{\rm CO}=1.8\times10^{20}$ cm$^{-2}$ (K km s$^{-1}$ pc$^2$)$^{-1}$ to estimate the H$_2$ column density from the CO integrated intensity \citep{Dame:2001lr}.
$\HI$ data were obtained with Very Large Array (VLA) as a part of The $\HI$ Nearby Galaxy Survey \citep[THINGS,][]{Walter:2008fr}.
The $\HI$ column density is estimated according to Equation~(5) of \cite{Walter:2008fr}.
The sensitivity of $\Sigma_{\rm gas}$ is 32~M$_\odot$ pc$^{-2}$ for NGC~3627, 11~M$_\odot$ pc$^{-2}$ for NGC~5055, and 30~M$_\odot$ pc$^{-2}$ for M~83.

The integrated intensity CO(3-2)/CO(1-0) ratio is defined as,
\begin{equation}
R_{31} = \frac{I_{\rm CO(3-2)}}{I_{\rm CO(1-0)}},
\end{equation}
where $I_{\rm CO(3-2)}$ and $I_{\rm CO(1-0)}$ are integrated intensities of CO(3-2) and CO(1-0), respectively.
CO(3-2) data were obtained with the Atacama Submillimeter Telescope Experiment (ASTE) telescope for M~83 \citep{Muraoka:2009vn} and with the James Clerk Maxwell Telescope (JCMT) for NGC~3627 and NGC~5055 as a part of the JCMT Nearby Galaxies Legacy Survey \citep[NGLS,][]{Wilson:2012fk}.

\subsection{SFR surface density}
\label{sec:sfr}

We combined H$\alpha$ and 24~$\mu$m data to estimate dust extinction-corrected H$\alpha$ luminosity using to Equation (5) of \citet{Calzetti:2007ys}\footnote{
The diffuse emission of H$_\alpha$ and 24~$\mu$m was not extracted, although it can also affect the slope of KS relation \citep{Liu:2011kx}.
The effect of the diffuse emission on SFR estimation and our results is discussed in Appendix.
}.
Then, SFR surface density ($\Sigma_{\rm SFR}$) is estimated from the H$\alpha$ luminosity according to Equation (2) of \cite{Kennicutt:1998fk}.
H$\alpha$ and 24~$\mu$m data are retrieved from the SIRTF Nearby Galaxies Survey \citep[SINGS,][]{Kennicutt:2003eu} for NGC~3627 and the Spitzer Local Volume Legacy \citep[LVL,][]{Dale:2009fv} for NGC~5055 and M~83, respectively.
[NII] contribution to H$\alpha$ luminosity was subtracted for all the data.

\section{Results}
\label{sec:result}

\subsection{Spatial distribution of CO(3-2)/CO(1-0)}
\label{sec:distribution}

\begin{figure*}
\begin{center}
\includegraphics[width=180mm]{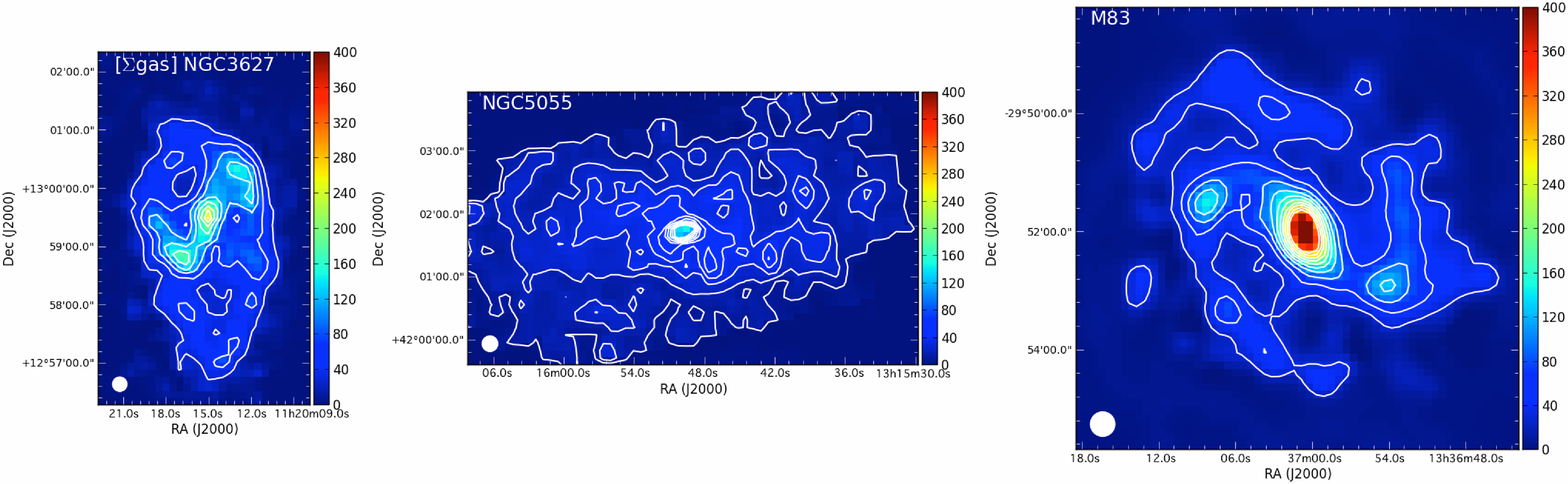}
 \caption{$\Sigma_{\rm gas}$ (M$_\odot$ pc$^{-2}$) maps of the sample galaxies.
 Intensity scales are set to be identical among the galaxies.
 The unit for the color bar is M$_\odot$ pc$^{-2}$.
 The spatial resolutions are shown on the bottom-left corner of each plot.
 The pixel sizes are set to 7.3~arcsec for NGC~3627, 7.0~arcsec for NGC5055 and 7.5~arcsec for M~83.
 White contours trace the distribution of $\Sigma_{\rm gas}$ with contour levels of $1\sigma$, $2\sigma$, $3\sigma$, $4\sigma$, $5\sigma$, $6\sigma$, $7\sigma$, $8\sigma$, $9\sigma$, $10\sigma$.
 $1\sigma$s for NGC~3627, NGC~5055, M~83 are 32, 11, and 30 M$_\odot$ pc$^{-2}$, respectively.}
\label{fig:sigmagas_sample}
\end{center}
\end{figure*}

\begin{figure*}
\begin{center}
\includegraphics[width=180mm]{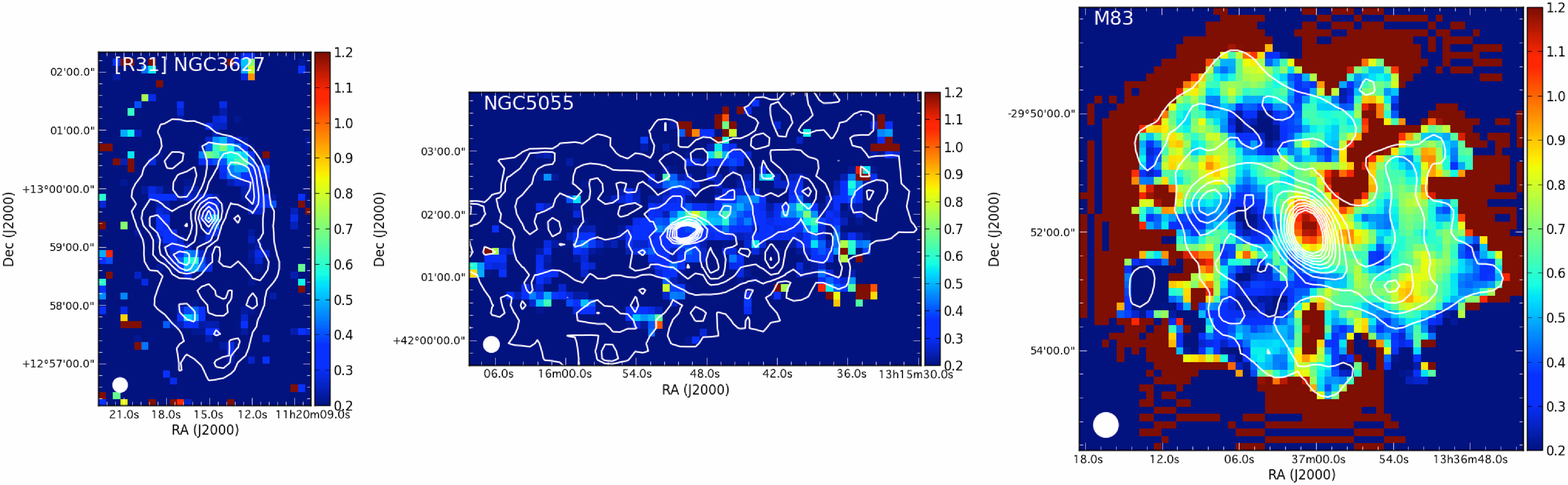}
 \caption{R$_{31}$ maps of the sample galaxies.
 Intensity scales are set to be identical among the galaxies.
 The spatial resolutions are shown on the bottom-left corner of each plot.
 The pixel sizes are set to 7.3~arcsec for NGC~3627, 7.0~arcsec for NGC5055 and 7.5~arcsec for M~83.
As in Figure~\ref{fig:sigmagas_sample}, $\Sigma_{\rm gas}$ distribution is shown as white contours as a reference.}
\label{fig:r31_sample}
\end{center}
\end{figure*}

\begin{figure*}
\begin{center}
\includegraphics[width=180mm]{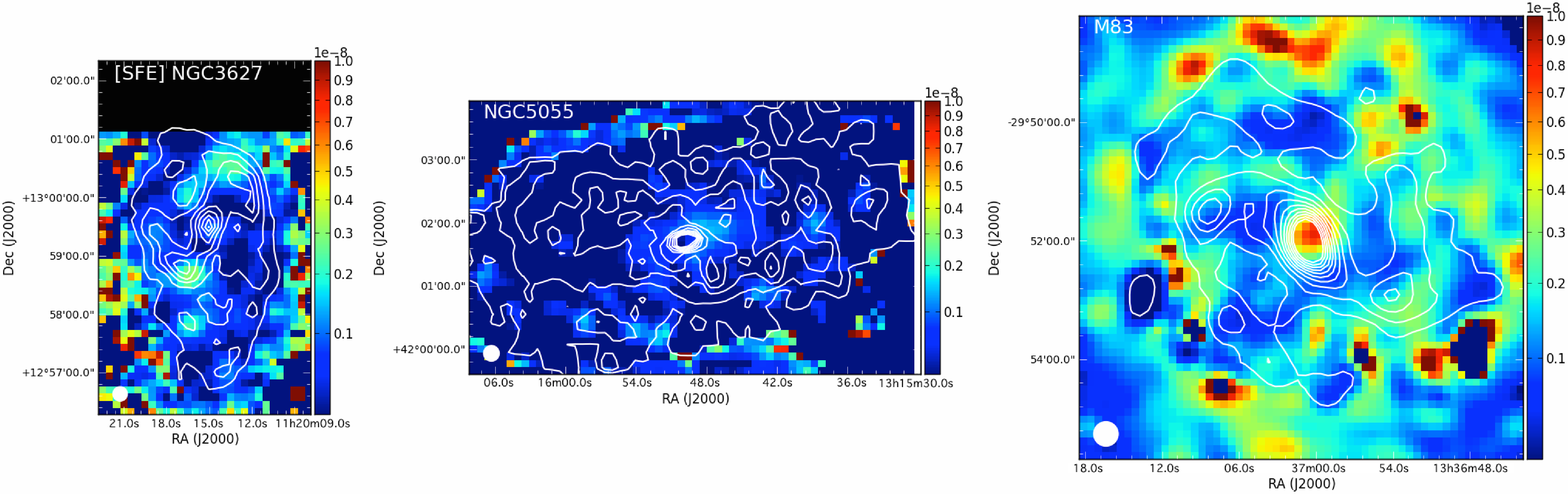}
 \caption{SFE (yr$^{-1}$) maps of the sample galaxies.
 Intensity scales are set to be identical among the galaxies.
 The unit for the color bar is yr$^{-1}$.
 The spatial resolutions are shown on the bottom-left corner of each plot.
 The pixel sizes are set to 7.3~arcsec for NGC~3627, 7.0~arcsec for NGC5055 and 7.5~arcsec for M~83.
 As in Figure~\ref{fig:sigmagas_sample}, $\Sigma_{\rm gas}$ distribution is shown as white contours as a reference.}
\label{fig:sfe_sample}
\end{center}
\end{figure*}

Figures~\ref{fig:sigmagas_sample}, \ref{fig:r31_sample} and \ref{fig:sfe_sample} respectively show $\Sigma_{\rm gas}$, $R_{31}$ and star formation efficiency (SFE $=\Sigma_{\rm SFR}/\Sigma_{\rm gas}$) maps of the sample galaxies.
The intensity scales for each figure were set to be the same.
The $\Sigma_{\rm gas}$ distributions are shown in all the maps as white contours as references.
The outermost contours in Figures~\ref{fig:sigmagas_sample}, \ref{fig:r31_sample} and \ref{fig:sfe_sample} correspond to the $\Sigma_{\rm gas}$ sensitivities of each galaxy.
$\Sigma_{\rm gas}$ is high at the central regions of all the sample galaxies and at the bar regions, especially the bar-end regions of NGC~3627 and M~83, as shown in Figure~\ref{fig:sigmagas_sample}.
The spiral arm structures can be also seen in the $\Sigma_{\rm gas}$ maps of NGC~3627 and M~83.
For the $R_{31}$ and SFE maps, the overall trends resemble each other: high values at downstreams of the bar-end regions and central regions.

\subsection{CO(3-2)/CO(1-0) ratio and the KS-relation}

\begin{figure*}
\begin{center}
\includegraphics[width=160mm]{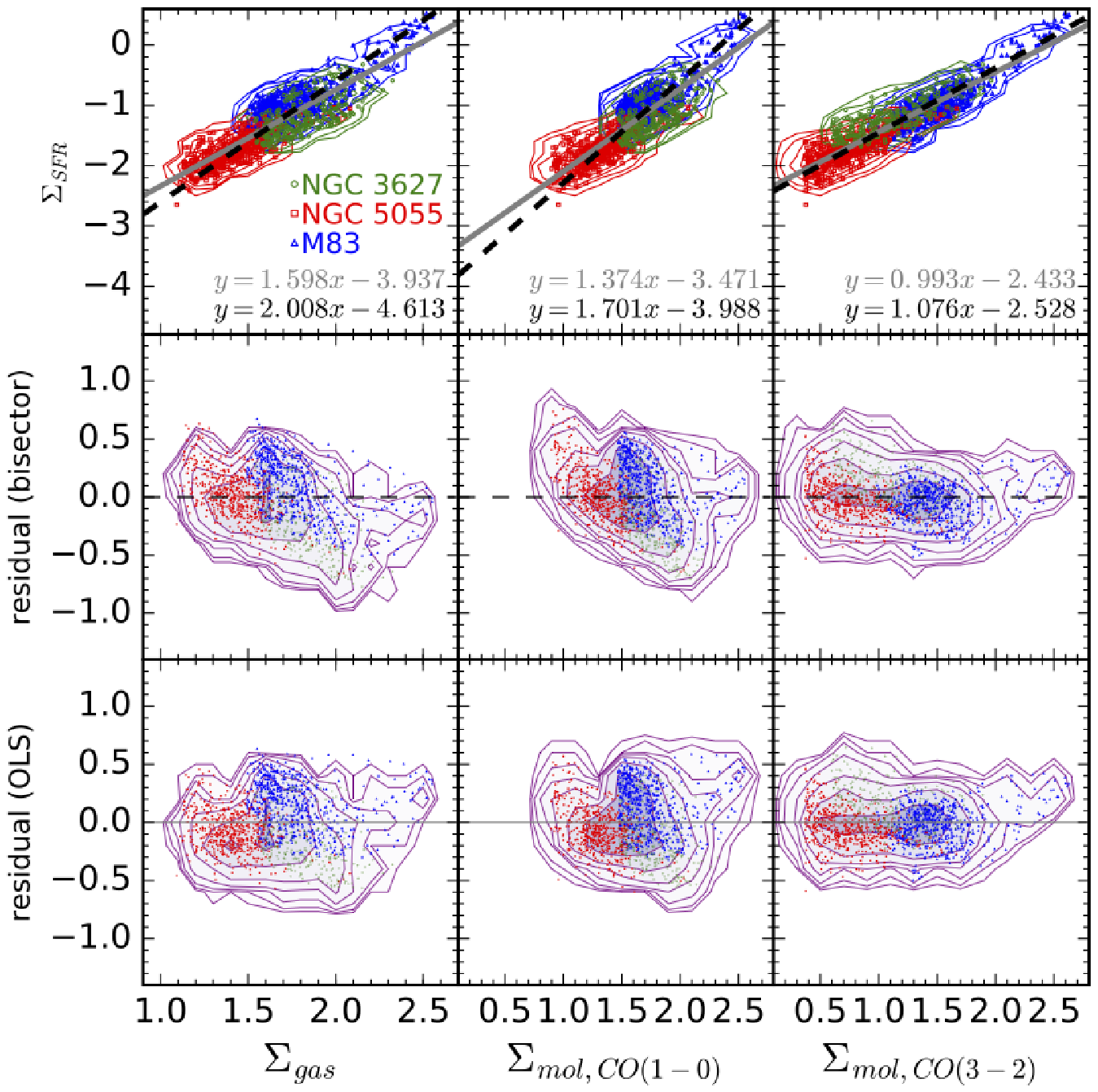}
 \caption{
[1st row] KS relation (left), CO(1-0)-based (center) and CO(3-2)-based (right) molecular KS relation plots with the observational data of NGC~3627 (green circle and contour), NGC~5055 (red square and contour) and M~83 (blue triangle and contour).
 Fitting results are shown at the bottom right corner of each panel, OLS($\Sigma_{\rm SFR} | \Sigma_{\rm gas}$) in grey (solid) and bisector in black (dash).
[2nd \& 3rd rows] Residual plots of bisector fitting (2nd row) and OLS($\Sigma_{\rm SFR}|\Sigma_{\rm gas}$) fitting (3rd row) for each KS relation.
 Contours show the density of all the data points for three galaxies.
 $\Sigma_{\rm mol, CO(3-2)}$ is inferred from CO(3-2) by assuming the same $X_{\rm CO}$ for CO(1-0) without CO(3-2)-to-CO(1-0) conversion.
 The qualitative trend in KS relation is similar to molecular KS relation since the galactic areas we studied are molecular-dominate region (H$_2/$\HI $>1$).
 The slope of CO(3-2)-based molecular KS relation is linear whereas the CO(1-0)-based one is super-linear, suggesting the difficulty in deriving molecular KS relation from CO(3-2) with a fixed CO(3-2)-to-CO(1-0) conversion factor.
 }
\label{fig:KS_obs8}
\end{center}
\end{figure*}

\begin{figure*}
\begin{center}
\includegraphics[width=180mm]{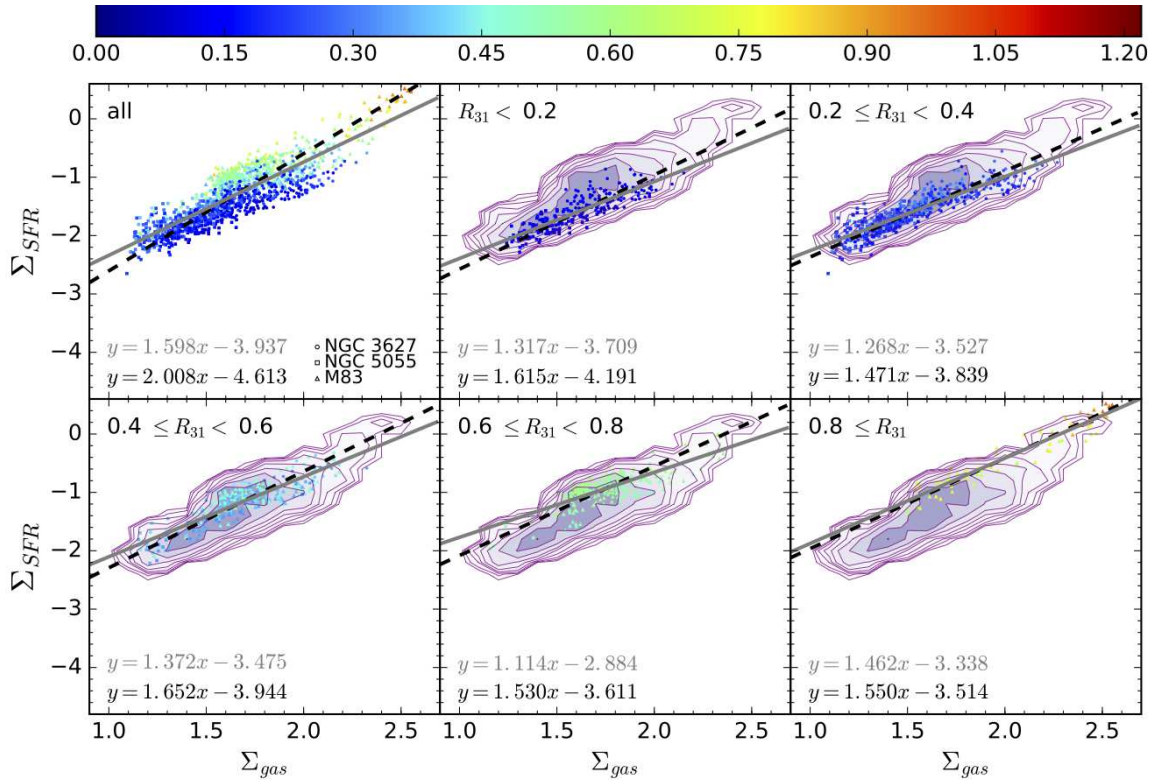}
 \caption{KS relation plots with the observational data of NGC~3627 (triangle), NGC~5055 (circle) and M~83 (square) color-coded for ranges of the integrated intensity ratio $R_{31}=\frac{\rm CO(3-2)}{\rm CO(1-0)}$.
 The contours show the density plots for whole sample as a reference. 
 Resultant KS-relation slopes are shown at the bottom right corner of each panel, OLS($\Sigma_{\rm SFR} | \Sigma_{\rm gas}$) in grey and bisector in black.
 The slopes for each subcategory with a fixed range of $R_{31}$ are shallower than the one for the whole sample.
 The normalizations for high $R_{31}$ subsamples tend to be high.
 }
\label{fig:KS_obs}
\end{center}
\end{figure*}

\begin{table*}
\begin{center}
\caption{Slopes and normalizations for each $R_{31}$ subsample \label{tab:kslawresult}}
\begin{tabular}{lcccc}
\tableline\tableline
& \multicolumn{2}{c}{OLS($\Sigma_{\rm SFR}|\Sigma_{\rm gas}$)} & \multicolumn{2}{c}{Bisector}\\
$R_{31}$ & Slope & Normalization & Slope & Normalization\\
\tableline
$<0.2$ & $1.32\pm0.05$ & $-(3.71\pm0.08)$ & $1.62\pm0.06$ & $-(4.19\pm0.09)$\\
$0.2-0.4$ & $1.27\pm0.03$ & $-(3.53\pm0.04)$ & $1.47\pm0.03$ & $-(3.84\pm0.05)$\\
$0.4-0.6$ & $1.37\pm0.04$ & $-(3.48\pm0.07)$ & $1.65\pm0.05$ & $-(3.94\pm0.08)$\\
$0.6-0.8$ & $1.11\pm0.05$ & $-(2.88\pm0.10)$ & $1.53\pm0.07$ & $-(3.61\pm0.12)$\\
$0.8-1.5$ & $1.46\pm0.05$ & $-(3.34\pm0.10)$ & $1.55\pm0.05$ & $-(3.51\pm0.10)$\\
\tableline
$0.2-1.5$ & $1.60\pm0.03$ & $-(3.94\pm0.04)$ & $2.01\pm0.03$ & $-(4.61\pm0.05)$\\
\tableline
\end{tabular}
\end{center}
\end{table*}

Figure~\ref{fig:KS_obs8} shows $\Sigma_{\rm SFR}-\Sigma_{\rm gas}$ relation (KS relation), CO(1-0)-based and CO(3-2)-based molecular KS relations: molecular gas surface densities, $\Sigma_{\rm mol, CO(1-0)}$ and $\Sigma_{\rm mol, CO(3-2)}$ are inferred from CO(1-0) and CO(3-2) integrated intensities, respectively.
$\Sigma_{\rm mol, CO(3-2)}$ is converted from CO(3-2) integrated intensity by assuming the same $X_{\rm CO}$ for CO(1-0) without CO(3-2)-to-CO(1-0) conversion.
The KS-relation slopes were derived using ordinary least-squares (OLS) regression of $\Sigma_{\rm SFR}$ on $\Sigma_{\rm gas}$, OLS($\Sigma_{\rm SFR}|\Sigma_{\rm gas}$), and the bisector of OLS($\Sigma_{\rm SFR}|\Sigma_{\rm gas}$) and OLS($\Sigma_{\rm gas}|\Sigma_{\rm SFR}$).
The slope of CO(3-2)-based molecular KS relation is linear whereas the CO(1-0)-based one is super-linear in both the fitting methods, suggesting the difficulty in deriving molecular KS relation from CO(3-2) with a fixed CO(3-2)-to-CO(1-0) conversion factor.
As previous studies have shown, the slope of the bisector of the two OLSs is steeper than that of OLS($\Sigma_{\rm SFR}|\Sigma_{\rm gas}$) \citep[e.g.,][]{Isobe:1990qy}.
A tighter relation for CO(3-2)-based molecular KS relation than CO(1-0)-based one indicates that CO(3-2) traces molecular gas that is more closely related to star formation.
We only used pixels for which CO(1-0), CO(3-2) and $\HI$ were detected.
Therefore, the area we studied here is somehow biased to molecular-dominated regions: all data points have H$_2/$\HI$>1$.

\begin{figure}
\begin{center}
\includegraphics[width=90mm]{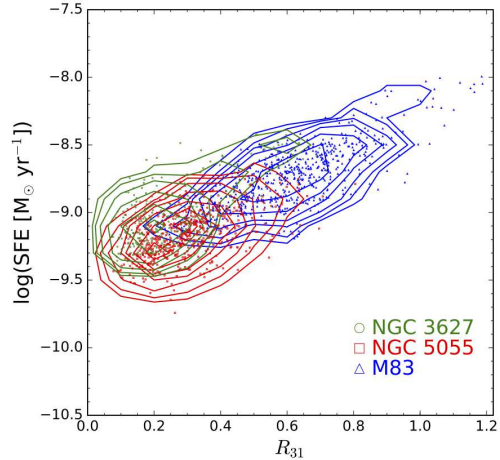}
 \caption{$R_{31}$ and SFE of NGC~3627 (green circle and contour), NGC~5055 (red square and contour) and M~83 (blue triangle and contour).
 $R_{31}$ is correlated with SFE over nearly 2 dex.}
\label{fig:KS_obs4}
\end{center}
\end{figure}

Figure~\ref{fig:KS_obs} shows the KS relation of NGC~3627, NGC~5055, and M~83.
KS relation plots are divided into subsamples based on the $R_{31}$ values -- these are color-coded according to $R_{31}$ in Figure~\ref{fig:KS_obs}.
The slopes of the KS relations for each sub-sample of $R_{31}$ are shallower than that obtained when fitting to all data concurrently.
The normalization for higher-$R_{31}$ subsamples tend to be higher than those for the lower-$R_{31}$ subsamples (see Table~\ref{tab:kslawresult}).
These trends are independent of the fitting methods: OLS($\Sigma_{\rm SFR} | \Sigma_{\rm gas}$) or a bisector of OLS($\Sigma_{\rm SFR} | \Sigma_{\rm gas}$) and OLS($\Sigma_{\rm gas} | \Sigma_{\rm SFR}$).
This trend is still seen even if we use $\Sigma_{\rm mol}$ instead of $\Sigma_{\rm gas}$.
We emphasize here that the different galaxies follow the same $\Sigma_{\rm SFR}$-$\Sigma_{\rm gas}$ relations for each $R_{31}$ even though data were not all obtained in the same conditions (observing seasons and instruments used).
This trend suggests that $R_{31}$ is correlated with the $\Sigma_{\rm SFR}$/$\Sigma_{\rm gas}$ ratio -- i.e., SFE.
Indeed, Figure~\ref{fig:KS_obs4} shows a clear correlation between $R_{31}$ and SFE.
In this plot, blue, green, and red symbols indicate M~83, NGC~3627, and NGC~5055, respectively.
This correlation has been already reported for nuclei of 60 nearby IR bright galaxies \citep{Yao:2003lr} and for M~83 data \citep{Muraoka:2007rt}.
We confirm that the correlation is seen even when adding the data of NGC~3627 and NGC~5055 to that of M~83.

\begin{figure}
\begin{center}
\includegraphics[width=90mm]{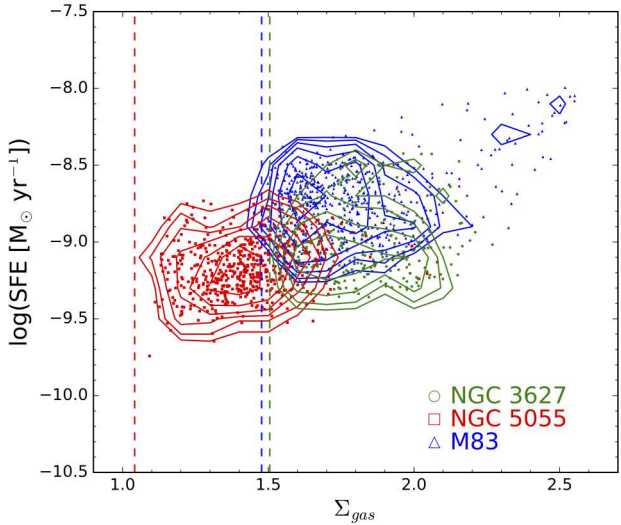}
 \caption{$\Sigma_{\rm gas}$ and SFE.
 Symbols are the same as Figure \ref{fig:KS_obs4}.
 Blue/red/green dashed lines represent the sensitivities of $\Sigma_{\rm gas}$ measurements for M~83 (30~M$_\odot$ pc$^{-2}$)/ NGC~5055 (11)/NGC~3627 (32).
 The boundaries of lower $\Sigma_{\rm gas}$ for each galaxy are determined by the sensitivities.
 The distributions are different from galaxy to galaxy.
 The weak but existing positive dependence of SFE on $\Sigma_{\rm gas}$ generates a super-linear KS relation.
}
\label{fig:KS_obs3}
\end{center}
\end{figure}

Figure~\ref{fig:KS_obs3} shows the relationship between SFE and $\Sigma_{\rm gas}$.
For each galaxy, there is no clear dependence of SFE on $\Sigma_{\rm gas}$, except for M~83: M~83 has outliers with high $\Sigma_{\rm gas}$ and SFE.
If the data for all the galaxies are combined, a weak $\Sigma_{\rm gas}$ dependence of SFE is observed, even when excluding the outlier data points with high $\Sigma_{\rm gas}$ and SFE values from M~83.
This explains the super-linear KS relation of our samples;
if there is no $\Sigma_{\rm gas}$ dependence of SFE among the samples, the resulting KS-relation slope is expected to be linear.

\begin{figure}
\begin{center}
\includegraphics[width=90mm]{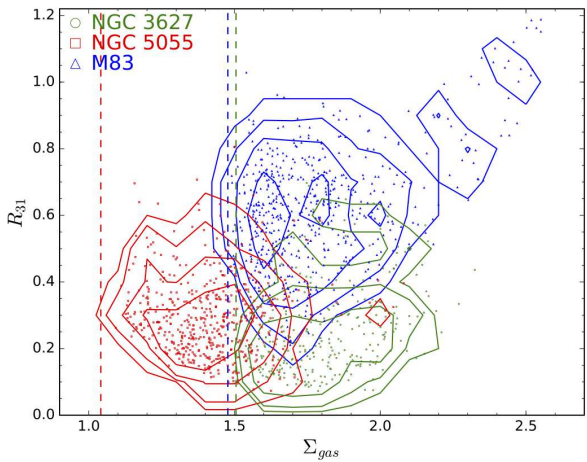}
 \caption{$\Sigma_{\rm gas}$ and $R_{31}$.
 Symbols are the same as Figure \ref{fig:KS_obs4}.
 Blue/red/green dashed lines represent the sensitivities of $\Sigma_{\rm gas}$ measurements for M~83 (30~M$_\odot$ pc$^{-2}$)/ NGC~5055 (11)/NGC~3627 (32).
 The trend is almost identical with Figure \ref{fig:KS_obs3}, since $R_{31}$ is correlated with SFE (see Figure \ref{fig:KS_obs4}).
 }
\label{fig:KS_obs5}
\end{center}
\end{figure}

\begin{figure}
\begin{center}
\includegraphics[width=90mm]{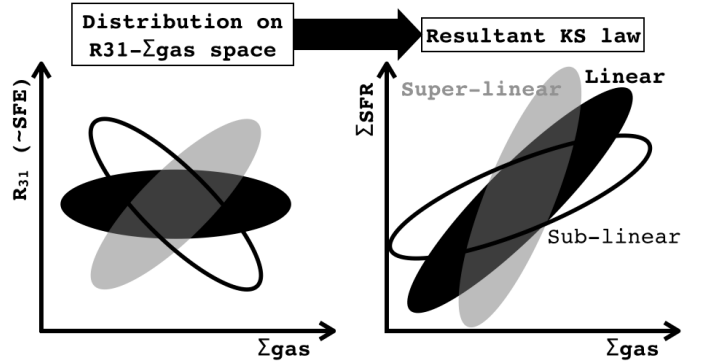}
 \caption{Relationship between the distribution of samples in the $R_{31}$-$\Sigma_{\rm gas}$ plane (left) and the resulting KS relation (right).
 The positive/no/negative correlations in R31-$\Sigma_{\rm gas}$ result in super-linear/linear/sub-linear KS-relation slopes.}
\label{fig:schematic}
\end{center}
\end{figure}

There is a similar trend in the $R_{\rm 31}$-$\Sigma_{\rm gas}$ relationship in Figure~\ref{fig:KS_obs5}.
This agrees with our expectations, since there exists a correlation between SFE and $R_{\rm 31}$ (Figure~\ref{fig:KS_obs4}).
Therefore, the distribution in the $R_{\rm 31}$-$\Sigma_{\rm gas}$ plane of the sample used to investigate the KS relation affects the slope of the resulting KS relation.
We summarize the relation between the location of sample galaxies in the $\Sigma_{\rm gas}$-$R_{31}$ plane and the resulting KS relation in Figure \ref{fig:schematic}.
If a bimodal distribution exists along the $R_{31}$ direction in $\Sigma_{\rm gas}$-$R_{31}$ space, one obtains a KS relation with two sequences.
M~83 shows a faint bimodal distribution in Figure~\ref{fig:KS_obs}: one from the majority (disk region in Figure~\ref{fig:r31_sample}) and the other from the outliers (nuclear starburst region) in $\Sigma_{\rm gas}$-$R_{31}$ space.
In Figure~\ref{fig:KS_obs5}, we also notice that our three sample galaxies distribute differently in the $R_{31}$-$\Sigma_{\rm gas}$ plane, even though they share similar stellar mass and SFR, --i.e., they are the same galaxy populations as a first-order approximation.
More specifically, M~83 has higher $R_{31}$ ratio than NGC~3627 while they share similar $\Sigma_{\rm gas}$, and on the other hand, NGC~3627 has a higher $\Sigma_{\rm gas}$ value than NGC~5055 while they share similar $R_{31}$.

\section{Hierarchical KS relation and rate-limiting step for star formation}
\label{sec:discussion}

In the previous section, we showed that the $R_{31}$-$\Sigma_{\rm gas}$ relation is different from galaxy to galaxy and its variety can impose the variety on the resultant KS relation.
In this section, we first define a ``hierarchical KS relation'', and consider our result in its context.
Then, we discuss a rate-limiting hierarchical ISM step that induces a variety in the KS relation of disk galaxies and the relationship with the galactic structures.

\subsection{Hierarchical KS relation}

There exists a hierarchy in the ISM within a galaxy \citep{Bergin:2007mz}:
cores that make up the material of stars in which the mass of the core correlates linearly with SFR (example tracer: HCN, mass of $0.5-5$~M$_\odot$, size of $0.03-0.2$~pc, and the mean density of $10^4-10^5$~cm$^{-3}$),
clumps ($^{13}$CO(1-0), $50-500$~M$_\odot$, $0.3-3$~pc, and $10^3-10^4$~cm$^{-3}$),
molecular clouds ($^{13}$CO(1-0) and $^{12}$CO(1-0), $10^3-10^4$ M$_\odot$, $2-15$~pc, $50-500$~cm$^{-3}$),
and diffuse atomic gas.
In addition, some studies also suggested the existence of substantial amounts of volume-filling molecules or diffuse small clouds traced by CO(1-0) in Galactic sources \citep{Polk:1988ly,Knapp:1988gf,Carpenter:1995qf,Heyer:1996pd,Hily-Blant:2007bh,Goldsmith:2008lq,Roman-Duval:2016mz} as well as extragalactic sources \citep{Young:1986ul,Wilson:1994ve,Rosolowsky:2007dq,Pety:2013rr,Caldu-Primo:2013la,Rahman:2011ys,Shetty:2014oq,Shetty:2014qe,Morokuma-Matsui:2015cr}.
Considering that the transitions between diffuse gas and self-gravitating gas, and between atomic gas and molecular gas depend differently on the ISM properties, the existence of non-self gravitating molecular gas traced by CO is not surprising \citep[e.g.,][]{Elmegreen:1993oq}.

The KS relation connects the amount of total gas and SFR of a galaxy or a certain region of galaxies.
Considering the hierarchy in the ISM, there are mainly four steps for star formation from gas: conversions from atomic gas to molecular gas, from molecular gas to molecular clouds, from molecular clouds to dense cores\footnote{Here, we do not distinguish clumps from clouds since they are relatively similar compared to cores in terms of size and mass.
Nowadays, the filaments found in molecular clouds are considered to correspond to clumps.
The $Herschel$ satellite revealed the ubiquity of filamentary structures \citep[e.g.,][]{Andre:2010wd} and it is shown that both cores for massive and low mass stars are formed in the $\sim0.1$ pc-wide filaments \citep{Andre:2016ly}.
The core mass fraction over filament and filament mass fraction over molecular clouds should be measured in future.}
, and from dense cores to stars.
We define the hierarchical KS relation as,
\begin{equation}
\Sigma_{\rm SFR} \propto f_{\rm core} f_{\rm SF, cloud} f_{\rm cloud} f_{\rm mol} \Sigma_{\rm gas}
\label{eq:hierarchicalKS},
\end{equation}
where $f_{\rm core} = \Sigma_{\rm core} / \Sigma_{\rm SF, cloud}$ (a ratio of surface densities of dense core, $\Sigma_{\rm core}$, within a cloud and its parent star-forming cloud, $\Sigma_{\rm SF, cloud}$, conversion efficiency from star-forming cloud to the dense core),
$f_{\rm SF cloud} = \Sigma_{\rm SF, cloud} / \Sigma_{\rm cloud}$ (a ratio of surface densities of the star-forming cloud and all the molecular clouds, $\Sigma_{\rm cloud}$),
$f_{\rm cloud} = \Sigma_{\rm cloud} / \Sigma_{\rm mol}$ (a ratio of surface densities of molecular clouds and molecular gas, $\Sigma_{\rm mol}$),
and $f_{\rm mol} = \Sigma_{\rm mol} / \Sigma_{\rm gas}$ (the ratio of surface densities of molecular gas and total gas, $\Sigma_{\rm gas}$).
The surface areas used to calculate surface densities are defined by the spatial resolution of the observations.

Equation~(\ref{eq:hierarchicalKS}) is derived by assuming a linear relation of $\Sigma_{\rm SFR}\propto\Sigma_{\rm core}$ claimed in the previous studies \citep[e.g.,][]{Solomon:1992kl,Gao:2004kx,Gao:2004fj,Wu:2005ai}\footnote{A sub-linear relation between $\Sigma_{\rm SFR}$ and $\Sigma_{\rm core}$ has been reported if limiting the sample to normal disk galaxies \citep{Usero:2015zr,Chen:2015lr,Bigiel:2016uq}.}.
No clear dependence of $f_{\rm core}$ on $\Sigma_{\rm SF, cloud}$ ($\sim$ CO(3-2) surface density) was reported based on low-mass star forming regions of our Galaxy, which supports a constant $f_{\rm core}$ (\citealt{Battisti:2014tg}, but see also \citealt{Meidt:2016fr}).
In addition, the linear relation of SFR-CO(3-2) seen in this study (Figure~\ref{fig:KS_obs8}, right) as well as previous studies \citep{Narayanan:2005qa,Komugi:2007xr,Bayet:2009mi,Iono:2009oz,Mao:2010kl,Greve:2014rw,Muraoka:2016lr} suggests $\Sigma_{\rm SFR}\propto\Sigma_{\rm SF, cloud}$, accordingly a constant $f_{\rm core}$.
It is important to investigate the conversion efficiency of each step and controlling factors for each efficiency, and then specify the rate-limiting step of star formation.

\subsection{ISM hierarchical step inducing variety in the KS relation}

We showed that SFE is correlated with $R_{31}$ in Figure~\ref{fig:KS_obs4}, suggesting that $R_{31}$ is one of the keys to determining star formation.
If we assume that CO(1-0) and CO(3-2) emissions respectively trace total molecular gas and star-forming molecular clouds\footnote{
The $f_{\rm mol}$ variety can impose the variety in the KS relation.
$f_{\rm mol}$ is shown to be higher for more massive galaxies than lower mass galaxies \citep[e.g.,][]{Bothwell:2014lr}, and higher at the smaller galactic radius than larger radius \cite[e.g.,][]{Schruba:2011zr,Tanaka:2014ys}.
Though in this study we focus on the relatively massive galaxies and their optical disks where $f_{\rm mol}$ is evenly high, $f_{\rm mol}$ may be a main contributor of the diversity of the KS relation in the atomic gas dominated environments, such as low mass galaxies and galaxy outskirts.
},
the $R_{\rm 31}$ value is a measure of a ratio of $\Sigma_{\rm SF, cloud}/\Sigma_{\rm mol}$, which corresponds to a product of $f_{\rm SF, cloud}$ and $f_{\rm cloud}$ in Equation~(\ref{eq:hierarchicalKS}).
At this point, it is not clear which ISM hierarchical step, formations of molecular cloud ($f_{\rm cloud}$) or star-forming molecular cloud ($f_{\rm SF, cloud}$), induces the $R_{31}$ variety since $\Sigma_{\rm SF, cloud}/\Sigma_{\rm mol}$ is a product of $f_{\rm cloud}$ and $f_{\rm SF, cloud}$.

We summarize the previous studies related to $f_{\rm cloud}$ and $f_{\rm SF, cloud}$ of galaxies in this section.
Note that there is no previous study on $f_{\rm cloud}$ and $f_{\rm SF, cloud}$ exactly in consideration of Equation~(\ref{eq:hierarchicalKS}).

$f_{\rm cloud}$ values for each galaxy were reported to be $\sim0.5$, however there is a variety as a function of galactic structure. 
\cite{Wilson:1994ve} reported $f_{\rm cloud}\sim0.4-0.7$ for M~33 based on a comparison of $^{12}$CO(1-0) and $^{13}$CO(1-0) emission.
For M~33, Rosolowsky et al.(2007) found a radial gradient of $f_{\rm cloud}$ in high-spatial resolution CO(1-0) maps: 0.6 at the galactic center and 0.2 at 4 kpc from the center.
\cite{Pety:2013rr} concluded that a half of the $^{12}$CO(1-0) flux of M~51 is from diffuse molecular gas suggested as a missing flux of the interferometric observations.
Assuming the same CO-to-H$_2$ conversion factor for molecular clouds and diffuse molecular gas, their study suggests $f_{\rm cloud}$ is expected to be $\sim0.5$.
For one of the early-type galaxies, NGC~4526, $f_{\rm cloud}$ is reported to be $\sim0.5$ \citep{Utomo:2015qy}.
In addition, using spatially resolved $^{12}$CO(1-0) and $^{13}$CO(1-0) data, higher $f_{\rm cloud}$ in the spiral arms than the inter-arm region is suggested in the Milky Way \citep{Sawada:2012fv,Sawada:2012fy} and NGC~3627 \citep{Morokuma-Matsui:2015cr}.

Although there is no direct measurements of $f_{\rm SF, cloud}$, the {\it number} fraction of star-forming molecular clouds over all the molecular clouds ($R_{\rm SF, cloud}=N_{\rm SF,cloud}/N_{\rm cloud}$, where $N_{\rm SF,cloud}$ and $N_{\rm cloud}$ are the numbers of star-forming and total clouds, respectively) has been investigated.
\cite{Kawamura:2009eu} classified molecular clouds in the Large Magellanic Cloud (LMC) based on the contiguity of clouds with $\HII$ regions and stellar clusters.
They found that $R_{\rm SF, cloud}$ is $\sim0.7$.
A similar value was also reported in M~33 \citep{Miura:2012qy}\footnote{\cite{Minamidani:2008wd} found that the $R_{31}$ is higher for molecular clouds with $\HII$ region and stellar clusters than the molecular clouds without them in LMC, and the same trend is confirmed in M~33 \citep{Miura:2012qy}.
This supports the idea that CO(3-2) traces the star-forming molecular clouds.}.

To convert $R_{\rm SF, cloud}$ to $f_{\rm SF, cloud}$, we must take into account the molecular cloud mass dependence of $R_{\rm SF, cloud}$ and the mass spectrum of molecular clouds, since there is a variety in the masses of molecular clouds.
\cite{Engargiola:2003fj} showed that massive molecular clouds tend to have higher $R_{\rm SF, cloud}$ values than low-mass molecular clouds.
For a variety in mass spectrum of molecular clouds of Local Group galaxies, \cite{Rosolowsky:2005lr} reported a steeper mass function (i.e., smaller contribution of large molecular clouds) for the outer disk of the Milky Way and M~33 compared to the inner disk of the Milky Way \citep[comparison of the mass spectra among Local Group galaxies is found in Figure~9 of][]{Blitz:2007qy}.
Recently, a steeper mass spectrum was reported in early-type galaxy, NGC~4526 \citep{Utomo:2015qy}.
In addition, it is shown that massive molecular clouds are likely to be found at the spiral arms in M~51 \citep{Koda:2009yq,Colombo:2014vn}.
These studies suggest that $f_{\rm SF, cloud}$ has a variety among galaxies, although the variation in the mass distributions has not been fully understood.

\subsection{What determines $R_{31}\sim f_{\rm SF, cloud} \times f_{\rm cloud}$ within a galaxy?}
\label{sec:structure}

In this section, we discuss what determines $f_{\rm cloud}$ and/or $f_{\rm SF, cloud}$ in galaxies.
A weak positive correlation seen in Figure~\ref{fig:KS_obs5} suggests that high $\Sigma_{\rm gas}$ is one of the necessary conditions for high $f_{\rm cloud}$ and/or $f_{\rm SF, cloud}$\footnote{
If $X_{\rm CO}$ is a function of $R_{31}$, the weak positive correlation observed in $R_{31}$-$\Sigma_{\rm gas}$ plot become steeper.
It is because the high $R_{31}$ gas is expected to have relatively high gas temperature \citep[e.g.,][]{Minamidani:2008wd}, accordingly smaller $X_{\rm CO}$.
This would work in the left direction in $R_{31}$-$\Sigma_{\rm gas}$ plot for high-$\Sigma_{\rm gas}$ and $R_{31}$ data points.
Therefore, this may increase the KS-relation slope but cannot make the super-linear KS-relation linear or sub-linear.
}.
The major differences between the high-$\Sigma_{\rm gas}$ galaxies (NGC~3627 and M~83) and low-$\Sigma_{\rm gas}$ galaxy (NGC~5055) in our sample is the existence of galactic-scale structures such as spiral arms and bars:
NGC~3627 and M~83 are classified as barred-spiral galaxies with grand design spirals whereas NGC~5055 is a spiral galaxy with flocculent spiral arms \citep{Elmegreen:1987kq}.
High gas density at the spiral arm, bar and central regions is seen in Figure~\ref{fig:sigmagas_sample} and reported in many observational studies \citep[e.g.,][]{Helfer:2003kx,Kuno:2007uq}.

Theoretical studies predict gas compression at spiral arms, i.e., the ``galactic shock'' in density wave theory \citep{Fujimoto:1968dw,Roberts:1969fc,Shu:1973ay} and a ``large-scale colliding flow'' in the dynamic spiral theory \citep{Dobbs:2008pt,Wada:2011kh,Baba:2017fk}.
A high gas density at the bar-end and leading-edge of the bar region is expected from the crowded orbits there \citep{Wada:1994mz}.
In addition, the bar structures are predicted to play a role in removing angular momentum, thus inducing gas inflow to the galactic center \cite[e.g.,][]{Wada:1992gf}.
Observations confirmed a higher gas density at the central regions of barred spiral galaxies compared to non-barred galaxies \citep{Sakamoto:1999rt,Kuno:2007uq}.

We can see that the downstream side of bar-end region have high $R_{31}$ and SFE in both NGC~3627 and M~83 in Figures~\ref{fig:r31_sample} and \ref{fig:sfe_sample}.
The offset between molecular gas and star formation at bar regions has been reported in \cite{Sheth:2002lr} based on interferometric observations.
They showed that H$\alpha$ emission is offset towards the downstream side of CO(1-0) not only at the bar-end but also within the bar, and the largest offsets are found in the strongest bars.
\cite{Kenney:1991fk} suggested that the offset star-forming regions in the downstream side of the bar-end in M~83 were presumably triggered as the gas passes through the transition regions of bar and spiral arms.
In addition, the enhancement of $R_{31}$ or SFE at the offset regions may be related to the smaller velocity dispersion of molecular gas there than bar-end region of NGC~3627 reported in \cite{Morokuma-Matsui:2015cr}.
The reduced random motion induces star formation.

Combined with the previous studies showing a high $f_{\rm cloud}$ and abundance of massive molecular clouds in spiral arms, our results suggest that galactic structures (dynamics) affect $\Sigma_{\rm gas}$ distribution, consequently $R_{31}\sim f_{\rm cloud} \times f_{\rm SF, cloud}$, and then star formation in galaxies.
It is theoretically predicted that gas compression increases not only $\Sigma_{\rm gas}$ but also $f_{\rm cloud}$ \citep{Inoue:2008ff,Inoue:2009ec}, and possibly $f_{\rm SF, cloud}$.
However, high $\Sigma_{\rm gas}$ is not a sufficient condition for high $f_{\rm cloud}$ and/or $f_{\rm SF, cloud}$.
In Figure~\ref{fig:KS_obs5}, we can see that M~83 has higher $R_{31}$ ratio than NGC~3627 while they share similar $\Sigma_{\rm gas}$.
To explore the hierarchical KS relation in more depth, we must increase the number of samples and spatial resolution and directly measure the $f_{\rm cloud}$ and $f_{\rm SF, cloud}$ using the newest-generation instruments such as the Atacama Large Millimeter /submillimeter Array (ALMA), and the Northern Extended Millimeter Array (NOEMA).

\section{Conclusions}
\label{sec:summary}

The $R_{31}$ dependence of the KS-relation slopes was investigated.
We combined the literature maps of $\HI$, CO(1-0), CO(3-2), H$\alpha$, 24~$\mu$m of three nearby spiral galaxies -- NGC~3627, NGC~5055, and M~83 -- and estimated their $\Sigma_{\rm gas}$ and $\Sigma_{\rm SFR}$.
The main results are as follows.
\begin{itemize}
\item The slopes for the KS relations of the samples subdivided by $R_{31}$ are shallower than that for all of the datasets combined. The normalizations for high-$R_{31}$ samples tend to be higher than those for the low-$R_{31}$ counterparts.

\item The KS-relation slopes change according to the distribution in $R_{31}$-$\Sigma_{\rm gas}$ space of the plotted samples; no $\Sigma_{\rm gas}$ dependence of $R_{31}$ results in a linear slope of the KS-relation whereas a positive correlation between $\Sigma_{\rm gas}$ and $R_{31}$ results in a super-linear slope of the KS-relation.

\item The $R_{31}$-$\Sigma_{\rm gas}$ distribution is different from galaxy to galaxy as well as within a galaxy.
High $\Sigma_{\rm gas}$ and $R_{31}$ are found at the central and the bar-end regions suggesting that galactic structures (dynamics) play a role in determining the galactic KS relation.

\end{itemize}

We also discuss the rate-limiting ISM hierarchical step of star formation by introducing the concept of a hierarchical KS relation.
Assuming a linear relationship between core mass and SFR, the KS relation can be re-written as Equation~(\ref{eq:hierarchicalKS}).
Since $R_{31}$ is a measure of star-forming cloud fraction among total clouds, -- i.e., a product of $f_{\rm SF, cloud}$ and $f_{\rm cloud}$, our result suggests that $f_{\rm SF, cloud}$ and/or $f_{\rm cloud}$ is the rate-limiting ISM hierarchy which imposes a diversity on the resultant KS relation.
It is still unclear which ISM hierarchy, $f_{\rm SF, cloud}$ or $f_{\rm cloud}$, is the main contributor imposing large variation in the KS relation both within a galaxy and among galaxies in relatively high-$\Sigma_{\rm gas}$ regimes.
High spatial resolution and high sensitivity observations in CO(1-0) to measure $f_{\rm cloud}$ and $f_{\rm SF, cloud}$ are required to reveal the origin of the KS-relation diversity.

\acknowledgments
We are grateful to anonymous referee, Dr.~Junichi Baba, Dr.~Tomoki Morokuma, Dr.~Shinya Komugi, Dr.~Robert C. Kennicutt, Dr.~Tsutomu Takeuchi, Dr.~Kouichiro Nakanishi, Dr.~Fumi Egusa and Dr.~Nario Kuno for enlightening comments and suggestions.
We thank Dr. Eric Feigelson for his constructive comments on the statistical aspects.
KM thanks Dr.~Ross Burns for English language editing.

\appendix
\section{Diffuse emission subtraction for SFR estimation}

\begin{figure*}
\begin{center}
\includegraphics[width=160mm]{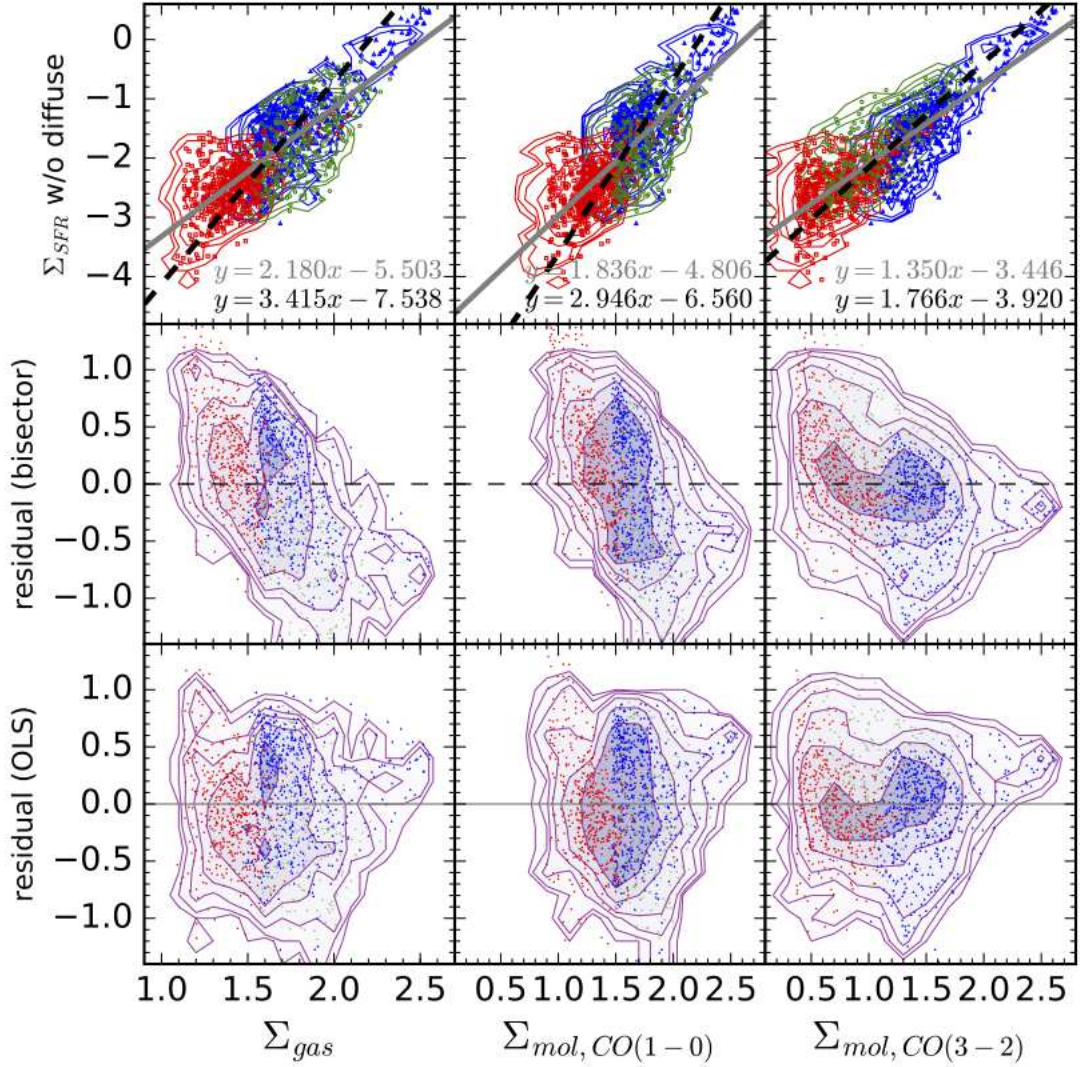}
 \caption{
Same figures as Figure~\ref{fig:KS_obs8} but with diffuse emission subtracted.
[1st row] KS relation (left), CO(1-0)-based (center) and CO(3-2)-based (right) molecular KS relation plots with the observational data of NGC~3627 (green circle and contour), NGC~5055 (red square and contour) and M~83 (blue triangle and contour).
 Fitting results are shown at the bottom right corner of each panel, OLS($\Sigma_{\rm SFR} | \Sigma_{\rm gas}$) in grey (solid) and bisector in black (dash).
[2nd \& 3rd rows] Residual plots of the bisector fitting (2nd row) and the OLS($\Sigma_{\rm SFR}|\Sigma_{\rm gas}$) fitting (3rd row) for each KS relation.
 Contours show the density of all the data points for the three galaxies.
 The slopes are steeper and the scatters are larger than those seen in Figure~\ref{fig:KS_obs8}.
 }
\label{fig:KS_obs8_wod}
\end{center}
\end{figure*}

\begin{figure*}
\begin{center}
\includegraphics[width=180mm]{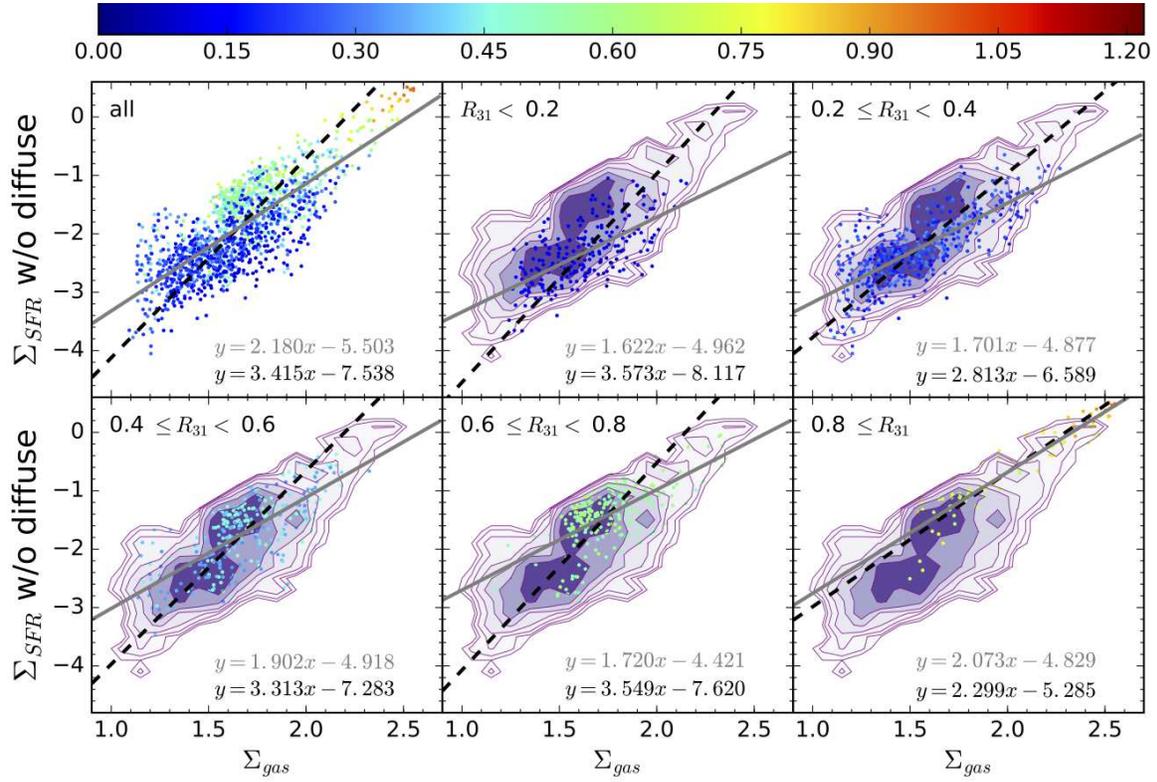}
 \caption{
Same figures as Figure~\ref{fig:KS_obs} but with diffuse emission subtracted data.
KS relation plots with the observational data of NGC~3627 (triangle), NGC~5055 (circle) and M~83 (square) color-coded for ranges of the integrated intensity ratio $R_{31}=\frac{\rm CO(3-2)}{\rm CO(1-0)}$.
 The contours show the density plots for whole sample as a reference. 
 Resultant KS-relation slopes are shown at the bottom right corner of each panel, OLS($\Sigma_{\rm SFR} | \Sigma_{\rm gas}$) in grey and bisector in black.
 The trends are similar to the ones seen in Figure~\ref{fig:KS_obs} where the slopes for each subcategory with a fixed range of $R_{31}$ tend to be shallower than the one for the whole sample and the normalizations for high $R_{31}$ subsamples tend to be high.
 }
\label{fig:KS_obs_wod}
\end{center}
\end{figure*}

\begin{table*}
\begin{center}
\caption{Slopes and normalizations for each $R_{31}$ subsample (w/o diffuse SFR) \label{tab:kslawresult_wod}}
\begin{tabular}{lcccc}
\tableline\tableline
& \multicolumn{2}{c}{OLS($\Sigma_{\rm SFR}|\Sigma_{\rm gas}$)} & \multicolumn{2}{c}{Bisector}\\
$R_{31}$ & Slope & Normalization & Slope & Normalization\\
\tableline
$<0.2$ & $1.62\pm0.13$ & $-(4.96\pm0.21)$ & $3.57\pm0.25$ & $-(8.12\pm0.41)$\\
$0.2-0.4$ & $1.70\pm0.07$ & $-(4.88\pm0.11)$ & $2.81\pm0.11$ & $-(6.59\pm0.16)$\\
$0.4-0.6$ & $1.90\pm0.10$ & $-(4.92\pm0.17)$ & $3.31\pm0.17$ & $-(7.28\pm0.28)$\\
$0.6-0.8$ & $1.72\pm0.12$ & $-(4.42\pm0.21)$ & $3.54\pm0.23$ & $-(7.62\pm0.41)$\\
$0.8-1.5$ & $2.07\pm0.09$ & $-(4.83\pm0.18)$ & $2.30\pm0.09$ & $-(5.28\pm0.19)$\\
\tableline
$0.2-1.5$ & $2.18\pm0.05$ & $-(5.50\pm0.08)$ & $3.41\pm0.07$ & $-(7.54\pm0.12)$\\
\tableline
\end{tabular}
\end{center}
\end{table*}

In section~\ref{sec:sfr}, $\Sigma_{\rm SFR}$ is estimated from H$\alpha$ and 24 $\mu$m maps of the galaxies under an assumption that all the H$\alpha$ emitting gas is ionized by the localized star formation in the same region.
However, the kpc-scale extra-planar ionized gas has been observed especially in edge-on galaxies, e.g., M~82 \citep[e.g.,][]{Lynds:1963rt}, indicating that ionizing radiation can reach farther than the sizes of an \HII~region of $\sim0.1-10$~pc in the Milky Way \citep[e.g.,][]{Kennicutt:1984fr,Garay:1999ys}.
The theoretical studies predict a larger escape fraction of ionizing radiation if the ISM is not smooth but structured \cite[][references therein]{Haffner:2009fk}.
Therefore, the H$\alpha$ emitting gas is not necessarily ionized by the localized star formation in the same region.

To distinguish a \HII~region from diffuse ionized gas and estimate $\Sigma_{\rm SFR}$ accurately, differences in spectroscopic property \citep{Madsen:2006kx,Blanc:2009vn} and spatial distribution \citep[e.g.,][]{Thilker:2000lr,Liu:2011kx} have been used.
In this study, we applied {\it HIIphot}, an IDL software developed by \cite{Thilker:2000lr}, to H$\alpha$ and 24~$\mu$m data to distinguish compact component from diffuse emission, following the procedures in \cite{Liu:2011kx}.

Figure~\ref{fig:KS_obs8_wod} is the same as Figure~\ref{fig:KS_obs8} but using $\Sigma_{\rm SFR}$ without diffuse emission.
As previous studies claimed \citep{Liu:2011kx,Momose:2013yq}, the resultant slopes become steeper since the contribution from diffuse emission is larger at lower $\Sigma_{\rm SFR}$ regime.
The increase in the scatter after subtraction of diffuse emission is also consistent with \cite{Liu:2011kx}.

Again, Figure~\ref{fig:KS_obs_wod} is the same as Figure~\ref{fig:KS_obs} but using $\Sigma_{\rm SFR}$ without diffuse emission.
In Figure~\ref{fig:KS_obs_wod} and Table~\ref{tab:kslawresult_wod}, we confirmed the same trends seen in Figure~\ref{fig:KS_obs} and Table~\ref{tab:kslawresult} where the slopes for the KS relations of the samples subdivided by $R_{31}$ are shallower than that for all the datasets combined and the normalizations for high-$R_{31}$ data tend to be higher than those for the low-$R_{31}$ counterparts.
Therefore, the contribution from diffuse ionized gas changes the slopes of the KS-relation but does not change the trends we obtained using $\Sigma_{\rm SFR}$ with diffuse H$\alpha$ and 24~$\mu$m emission.

\section{Akaike's Information Criterion for $\Sigma_{\rm gas}-\Sigma_{\rm SFR}$ relation}

\begin{figure*}
\begin{center}
\includegraphics[width=190mm]{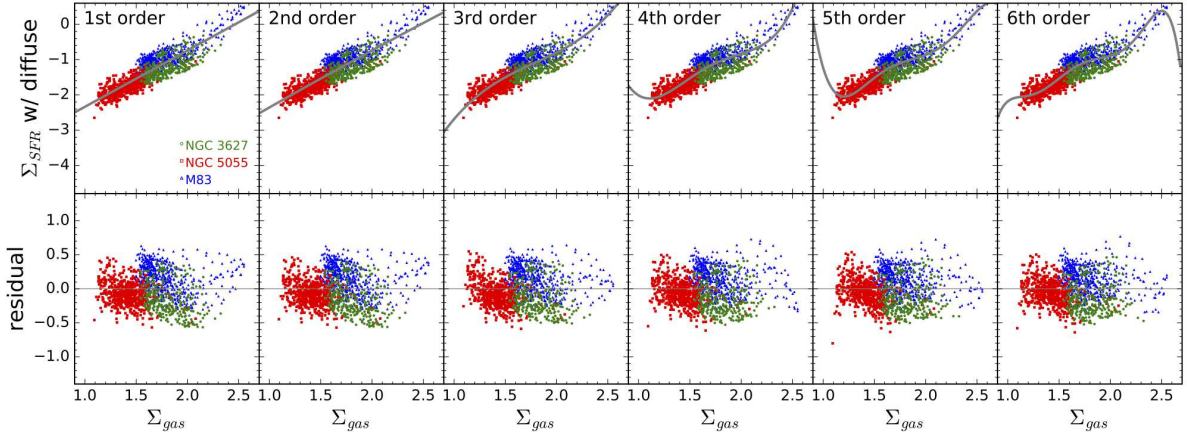}
 \caption{
  Result of the polynomial fitting (grey solid line) to the KS relation with the observational data of NGC~3627 (green triangle), NGC~5055 (red circle) and M~83 (blue square).
 Diffuse emission is not subtracted from $\Sigma_{\rm SFR}$.
 The residual plots are also presented in the bottom low.
 The AIC for each fitting is shown in Table~\ref{tab:AIC}.
 }
\label{fig:akaike}
\end{center}
\end{figure*}

\begin{figure*}
\begin{center}
\includegraphics[width=190mm]{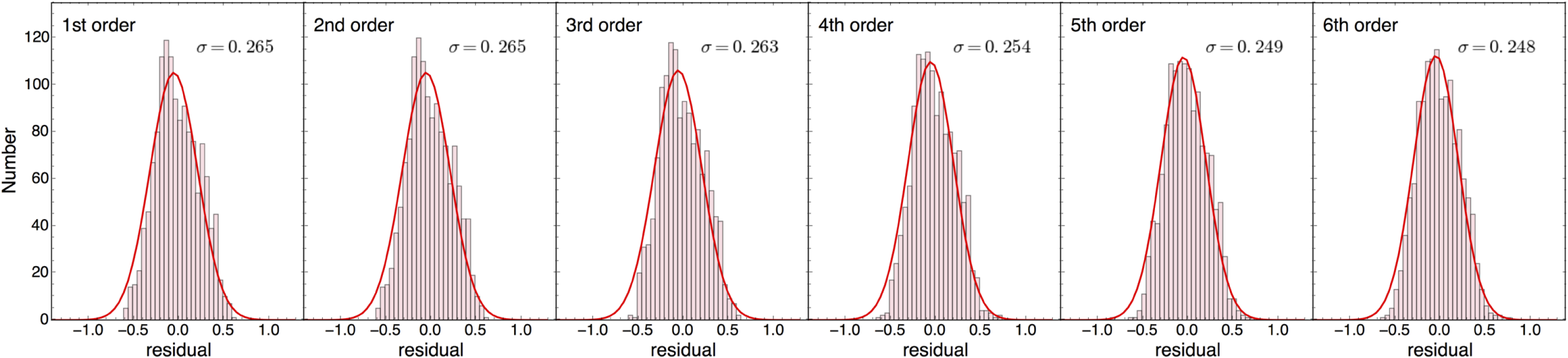}
 \caption{
  Residual distribution of Figure~\ref{fig:akaike}.
  The red solid lines show the fitting results with Gaussian whose $\sigma$ is shown at upper right corner of each plot.
 }
\label{fig:akaike_distribution}
\end{center}
\end{figure*}

\begin{figure*}
\begin{center}
\includegraphics[width=190mm]{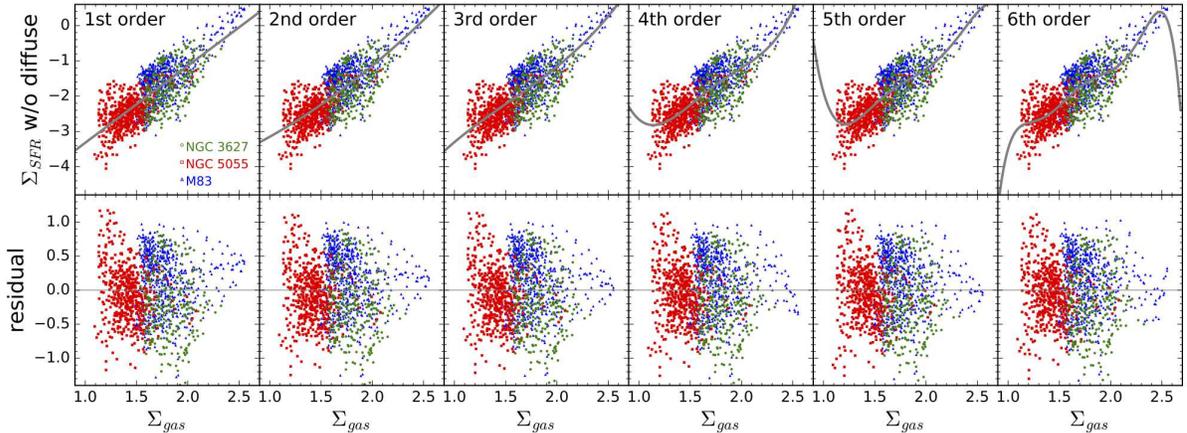}
 \caption{
 Same as Figure~\ref{fig:akaike} but using $\Sigma_{\rm SFR}$ without diffuse H$\alpha$ and 24~$\mu$m emission.
 }
\label{fig:akaike_wod}
\end{center}
\end{figure*}

\begin{figure*}
\begin{center}
\includegraphics[width=190mm]{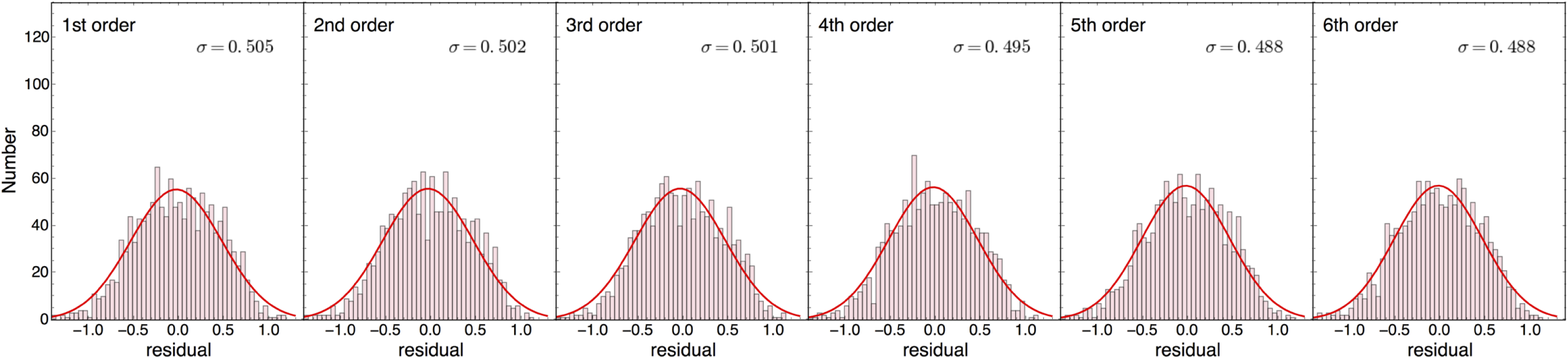}
 \caption{
  Residual distribution of Figure~\ref{fig:akaike_wod}.
  The red solid lines show the fitting results with Gaussian whose $\sigma$ is shown at upper right corner of each plot.
 }
\label{fig:akaike_distribution_wod}
\end{center}
\end{figure*}

\begin{table*}
\begin{center}
\caption{AIC of the polynomial fitting model for KS relation \label{tab:AIC}}
\begin{tabular}{lcccccc}
\tableline\tableline
& 1st & 2nd & 3rd & 4th & 5th & 6th\\
\tableline
w/ diffuse & 4.478 & 4.095 & -42.12 & -105.5 & -138.2 & \underline{-155.0}\\
w/o diffuse & 1734 & 1724 & 1721 & 1705 & 1697 & \underline{1684}\\
\tableline
\end{tabular}
\tablecomments{
The minima are shown with underline.
}
\end{center}
\end{table*}

We measure Akaike's Information Criterion \citep[AIC,][]{Akaike:1974yq} to test the validity of applying a linear fitting to the KS relation.
In general, fitting with a higher order polynomial function to data performs better than that of lower degree.
However, such high-order fitting is able to be {\it forced} to fit to the fluctuations that is not related to the true relation (``Overfitting'').
The AIC can deal with trade-off between the goodness of fit of the model and the complexity of the model.
The AIC is defined as
\begin{equation}
{\rm AIC} = -2 [ \mathscr{L} (\hat{\theta}) - K],
\end{equation}
where $\mathscr{L}$ is a likelihood function, $\hat{\theta}$ is a set of maximum-likelihood estimators, and $K$ is the number of free parameters of the assumed model.
The most preferred model minimizes the AIC.

Hereafter, we follow the Appendix of \cite{Takeuchi:2000vn} to calculate the AIC.
We adopted the $m$th order polynomial model to fit a set of $n$ pairs of observations, ($x_1, y_1$), ..., ($x_n, y_n$) as,
\begin{equation}
y_i = \sum_{l=0}^{m} a_l x_i^l + \epsilon_i,
\end{equation}
where we assume that $\epsilon_i$ is an independent random variable that follows the normal distribution with mean of zero and dispersion of $\sigma^2$.
In case of the normal distribution, the AIC is written as
\begin{equation}
{\rm AIC}(m) = n (\ln{2\pi} + 1) + n \ln{\widehat{\sigma^2}(m)} + 2 (m+2),
\label{eq:aic}
\end{equation}
where
\begin{equation}
\widehat{\sigma^2}=-\frac{1}{n}\sum^n_{i=1} \left( y_i - \sum^m_{l=0} \hat{a_l} x_i^l \right)^2.
\end{equation}
We calculate the AIC for $m=1\sim6$ using Equation~\ref{eq:aic}.

Figures~\ref{fig:akaike} and \ref{fig:akaike_wod} show the polynomial fitting results of the KS relation with and without diffuse emission for SFR estimation, respectively.
Figures~\ref{fig:akaike_distribution} and \ref{fig:akaike_distribution_wod} are the residual plots for the 1st$\sim$6th polynomial fittings in Figures~\ref{fig:akaike} and \ref{fig:akaike_wod}, respectively.
According to the AIC, $6$th-order polynomial is the most preferable for KS relation of our three target galaxies rather than linear function (see Table~\ref{tab:AIC}).
However, as we showed in Section~\ref{sec:result} and previous studies, the resultant KS relation can change according to what kind of data (galaxies, regions in a galaxy and resolution) are used to plot.

\bibliographystyle{apj}
\bibliography{apj-jour,/Users/kmatsui/Documents/Papers/myref_moro}

\end{document}